\documentclass[preprint,aps,prc,showpacs,nofootinbib]{revtex4}
\usepackage{epsfig}
\usepackage{graphicx,amsmath}

\newcommand\nn{\nonumber}

\def\phi{\varphi}

\newcommand\ba{\begin{eqnarray}}
\newcommand\ea{\end{eqnarray}}

\begin{document}

\title{$2\gamma$ and $3\gamma$  annihilation as  calibration processes for high energy $e^+ e^-$ colliders}

\author{E.~Barto\v{s}}
\affiliation{Institute of Physics, Slovak Academy of Sciences,
Bratislava}

\author{S. Bakmaev}
\affiliation{Joint Institute for Nuclear Research, Dubna, Russia}

\author{E.~A.~Kuraev}
\affiliation{Joint Institute for Nuclear Research, Dubna, Russia}

\author{M.~G.~Shatnev}
\affiliation{NSC KIPT, Kharkov, Ukraine}

\author{M.~Se\v{c}ansk\'{y}}

\affiliation{Joint Institute for Nuclear Research, Dubna, Russia}
\affiliation{Institute of Physics, Slovak Academy of Sciences,
Bratislava}

\date{\today}

\begin{abstract}
Born differential cross sections and the lowest-order radiative
correction to them are obtained in the kinematics of large-angle final
photons emission in high-energy electron-positron annihilation
processes. Taking into account possible emission of real soft
and hard photons in collinear kinematics we show the validity of the
Drell-Yan form of differential cross section with the nonsinglet
structure functions of initial leptons. The leading and next-to leading
contribution to the cross sections is obtained.
The relevant numerical estimations are presented.
\end{abstract}

\pacs{12.12.-m, 13.40.-f}
\maketitle

\section{Introduction}
The process of  Compton scattering,namely the $2\gamma$ ($3\gamma$) annihilation at high energy
collisions of electron and positron, which is under consideration here,
$$e^{-}(p_-)+e^{+}(p_+) \to \gamma(k_1) + \gamma(k_2), s=(p_-+p_+)^2
\gg m^2_e=m^2;$$ $$e^{-}(p_-)+e^{+}(p_+) \to \gamma(k_1) +
\gamma(k_2)+ \gamma(k_3),$$
is studied,
including radiative corrections (RC). This process was first considered in the famous
papers by L. Brown and R. Feynman and H.Harris and L. Brown in the early 1950-s and then
revised in 1973 by H. Berends and
R. Gastmans  \cite{Feynman}.

Nowadays these processes are used
in the collider physics as normalization processes which provide
an independent way to measure luminosity of beams. The
large-angle emission of final photons provides a clean signal
for independent method of luminosity measuring. The precise
knowledge of this process must also be taken into account when estimating
the background in channels with neutral meson production. These
reasons are the motivation of our paper.

Differential cross section of two-gamma channel in Born approximation
has the form
\ba
\frac{d\sigma_B}{dO_1}=\frac{\alpha^2}{s\beta}[\frac{1+\beta^2c^2}{1-\beta^2c^2}+
2\beta^2(1-\beta^2)\frac{1-c^2}{(1-\beta^2c^2)^2}],
\ea
with $\beta=\sqrt{1-\frac{4m^2}{s}}$, $c=\cos\theta$ and $\theta$ being the
polar angle between the initial electron and photon (with momentum $k_1$
 in the center of mass reference frame (which is implied below).
In the high-energy limit for large-angle photon emission we can
put $\beta=1$ in (1) and obtain
$$
\frac{d\sigma_B}{dO_1}=\frac{\alpha^2(1+c^2)}{s(1-c^2)}.
$$

The conservation law, on-mass shell conditions and
kinematic invariants are defined for the $e\bar{e} \to \gamma\gamma$
process as
\begin{eqnarray}
p_+ + p_- = k_1 + k_2, p_+^2=p_-^2= m^2, k_1^2=k_2^2=0, \nn \\
\chi_i=2p_-k_i, i=1,2, \chi_1+\chi_2=s.
\end{eqnarray}

The corresponding "total" cross section estimated for the region $\chi_1
\sim \chi_2 \sim s$  has an order
of $\frac{\pi \alpha^2}{s}$, which is rather large
compared to processes with weak and strong interactions included.
So their knowledge with RC of higher order is urgent since these
pure QED processes provide large background to study manifestation
of other than QED interactions.

Using annihilation processes with the aim  of calibration of beams
has an essential advantage compared to ones based on Bhabha
scattering and annihilation to a pair of charged particles $e\bar{e}
\to \mu \bar{\mu}, \tau \bar{\tau}$. Really, the process of $2\gamma$
annihilation at large angles has the same order as Bhabha one but
is free from difficulties connected with the final state interactions
with taking into account the vacuum polarization.

The theorem on factorization \cite{Lipatov} of hard and soft momenta in the cross section
of exclusive form permits one to take RC in the leading logarithmic
approximation
\begin{eqnarray}
\frac{\alpha}{\pi} \ll 1 , \quad \frac{\alpha}{\pi}L \sim 1, \quad
L=\log \frac{s}{m^2},
\end{eqnarray}
in all orders of perturbation theory in terms of structure functions
of electron and positron $D(x,L)$. This fact permits one to write
down the differential cross section of two quantum annihilation
\begin{eqnarray}
d\sigma(p_-,p_+,k_1,k_2)=\int_0^1 dx D(x,L) \int_0^1 dy
D(y,L)d\sigma_B(xp_-,yp_+,k_1,k_2)(1+\frac{\alpha}{\pi}K_2), \label{4}
\end{eqnarray}
with the "shifted" Born cross section of the form:
\ba
d\sigma_B(xp_-,yp_+,k_1,k_2)=\frac{2\alpha^2}{s x y}\frac{x^2(1-c)^2+y^2(1+c)^2}{[x(1-c)+y(1+c)]^2}d O_1.
\label{5}
\ea

Main attention must be paid to calculation of  $K$-factor $K=1+(\alpha/\pi)K_2$;
its knowledge permits
one to increase the accuracy of theoretical description up to $10^{-3}$ level.

\section{Virtual, soft real, and hard collinear photon emission contribution}

Using the known results of calculation of virtual corrections \cite{Feynman},
we obtain:
\begin{eqnarray}
\frac{d\sigma_{virt}}{d\sigma_B}=1+\delta_V=1+\frac{\alpha}{\pi} \Big[-
\frac{1}{2}L^2 -(L-1)\ln\frac{m^2}{\lambda^2}+\frac{3}{2}(L-1)+K_V \Big].
\end{eqnarray}
The explicit expression for the value $\delta_V$ in our kinematic region is given in Appendix.

\begin{figure}[t]
\begin{center}
\includegraphics{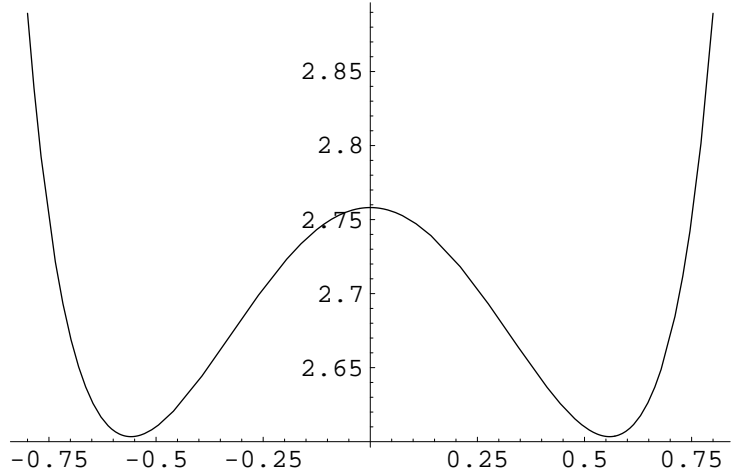}
\caption{The angular dependence of $K_{SV}(c)$ (see (\ref{Ksv})).}
\label{FigKsv}
\end{center}
\end{figure}

Emission of an additional soft photon avoiding the detectors (its energy in the c. m. frame is implied below
does not exceed some small quantity.The contribution for $\Delta E<< E=\sqrt{s}/2$ is:
\ba
\frac{d\sigma_{soft}}{d\sigma_B}=\delta_S=-\frac{\alpha}{4\pi^2} \int
\frac{d^3k}{\omega}(\frac{p_-}{p_-k}-\frac{p_+}{p_+k})^2\Big|_{\omega
< \Delta E} = \nn \\
\frac{\alpha}{\pi} \Big[
-\frac{1}{2}L^2 +L+2(L-1)\ln(\frac{2\Delta E }{\lambda})+K_S
\Big],
\ea
with $K_S=-\pi^2/3$.
 The total sum of the virtual and real soft photon emission has the form:
 \ba
 \frac{d\sigma_{virt}}{d\sigma_B}+\frac{d\sigma_{soft}}{d\sigma_B}=\frac{\alpha}{\pi}[(L-1)[2\ln\frac{\Delta E}{E}+\frac{3}{2}]+
 K_{SV}], \label{8}
 \ea
with
\ba
K_{SV}=\frac{\pi^2}{3}+\frac{1}{4(1+c^2)}[(5-6c+c^2)\ln\frac{1+c}{2}+ \nn \\
(5+2c+c^2)\ln^2\frac{1+c}{2}+(5+6c+c^2)\ln\frac{1-c}{2}+ \nn \\
(5-2c+c^2)\ln^2\frac{1-c}{2}]. \label{Ksv}
\ea
In Fig. \ref{FigKsv} we show the dependence of $K_{SV}(c)$.

A contribution from emission of a hard photon in collinear kinematics
can be obtained using the method of quasi-real electrons \cite{Baier}.
For this purpose we introduce some numerically small angle $\theta_0$ and consider emission
of an additional hard photon inside the cones $\theta<\theta_0$ along the electron and positron
direction of motion. The chiral amplitude method cannot be applied to this kinematical region
since the chirality is not a good quantum number for collinear kinematics.
We distinguish emission along the initial electron $\theta < \theta_0$, where $\theta$ is the angle between
3-momenta of electron (we choose this direction as the $z$ axis) and photon with energy $\omega$
\begin{eqnarray}
d\sigma^{coll}_{e^-}(p_-,p_+)=\frac{\alpha}{2\pi}
\int_0^{1-\frac{\Delta E}{E}} dx \Big[ \frac{1+x^2}{1-x}(L_\theta-1)+
1-x\Big]d\sigma_B(xp_-,p_+), x=1-\frac{\omega}{E},
\end{eqnarray}
and emission along positron: $\pi-\theta<\theta_0$ with numerically small value of $\theta_0<<1$
\begin{eqnarray}
d\sigma^{coll}_{e^+}(p_-,p_+)=\frac{\alpha}{2\pi}
\int_0^{1-\frac{\Delta E}{E}} dx \Big[ \frac{1+x^2}{1-x}(L_\theta-1)+
1-x\Big]d\sigma_B(p_-,xp_+),
\end{eqnarray}
with $L_\theta=L+\log(\frac{\theta_0}{2})^2,\theta_0 \ll 1$ and the expressions for the shifted
cross sections are given in (\ref{5}).

\section{Hard photon emission correction. Form of $K^{hard}$}

The contribution of  the kinematics when all three hard photons are  emitted in so called non-collinear
kinematics (all three photons are emitted outside cones $\theta < \theta_0, \quad \pi-\theta \ll
\theta_0$) can be obtained using the chiral amplitude method \cite{Eidelman}
\ba
d\sigma^{hard}=\frac{16\alpha^3}{3\pi^2s}R d\Phi,
\ea
with\footnote{In paper\cite{ihep} (1997) in the right hand sides of 4.3 and 4.6
the multipliers $\frac{1}{3!}$ and $\frac{1}{2}$ were lost.}
\ba
R=\frac{\nu_3^2(1+c_3^2)}{\nu_1^2\nu_2^2(1-c_1^2)(1-c_2^2)}+\frac{\nu_2^2(1+c_2^2)}{\nu_1^2\nu_3^2(1-c_1^2)(1-c_3^2)}+\nn \\
\frac{\nu_1^2(1+c_1^2)}{\nu_3^2\nu_2^2(1-c_3^2)(1-c_2^2)},
\ea
and
\ba
d\Phi=\frac{1}{s}\frac{d^3q_1}{2\omega_1}\frac{d^3q_2}{2\omega_2}\frac{d^3q_3}{2\omega_3}\delta^4(p_-+p_+-k_1-k_2-k_3),
\ea
$c_1,c_2,c_3$ are the cosines of the photon emission angles to the initial electron 3-momentum.
Using the on mass shell conditions and the conservation of 4-momentum law we can put the expression for $d\Phi$ in
one of the equivalent forms
\ba
d\Phi=\frac{1}{16}\frac{1-\nu_1}{[2-\nu_1(1-c_{13})]^2}\nu_1d\nu_1d O_1 d O_3= \nn \\
\frac{1}{16}\frac{1-\nu_3}{[2-\nu_3(1-c_{13})]^2}\nu_3d\nu_3d O_1 d O_3\quad...,
\ea
and similar ones obtained from this expression by permutations of the indices. Here $d O_1,d O_3$ are the
angular phase volumes and $c_{13}=1-2(1-\nu_2)/(\nu_1\nu_3)$ is the cosine of the angle between the
photons 1, 3 directions.

It can be seen that the dependence on the auxiliary parameters $(\Delta E/E)$ and $\theta_0$
will be cancelled in the expression of $K^{hard}$ defined as
\ba
\frac{\alpha}{\pi}d\sigma_B(xp_-,yp_+;k_1,k_2)K^{hard}=\int d\sigma^{hard}\Theta+ \nn \\ \frac{\alpha}{2\pi}\int\limits_0^{1-\frac{\Delta E}{E}}
\frac{dx}{1-x}[(1+x^2)\ln\frac{\theta_0^2}{4}+(1-x)^2][d\sigma(xp_-,p_+)+d\sigma(p_-,xp_+)].
\ea
Here the symbol $\Theta$ means restrictions on the manifold of integration variables $d\Phi$. They are:
all three energy fractions must exceed $\Delta E/E$ (hardness condition) and, besides, the conservation law
restrict $\vec{k}_1+\vec{k}_2+\vec{k}_3=0$. In particular,
\ba
c_1\nu_1+c_2\nu_2+c_3\nu_3=0;   \nu_1+\nu_2+\nu_3=2, \nu_i=\frac{\omega_i}{E}, \frac{\Delta E}{E}<\nu_i<1.
\ea
The noncollinear kinematics conditions must also be put on:
$\theta_0<\theta_i<\pi-\theta_0$.
Moreover, the experimental cuts connected with detection of the final photons can be included in this set of cuts.
These conditions depend on experimental set-ups.
It is them reason why we do not consider it in numerical estimations here.

Terms containing "large logarithm" $L$ can be put in the form
\begin{eqnarray}
\frac{\alpha}{2\pi}(L-1)[ \int_0^1 dx P^{(1)}(x) d\sigma_B(xp_-,p_+)
+ \int_0^1 dx P^{(1)}(x)d\sigma_B(p_-,xp_+)].
\end{eqnarray}
with
\begin{eqnarray}
P(x) = P^{(1)}(x) =\lim_{\sigma \to 0} [ \delta(1-x)(2\ln \sigma +
\frac{3}{2}) + \frac{1+x^2}{1-x}\theta(1-x-\sigma)], \  \int_0^1 dx
P^{(1)}(x)=0,
\end{eqnarray}
which is the kernel of the evolution equation of twist-2 operators [2]. The
parameter $x$ can be
interpreted as the energy function of electron considered as a
parton in the initial electron. In such a way, one can obtain the
general form of cross section in the form of cross-section of Drell-Yan
process given above (see (\ref{4})) with
\begin{eqnarray}
D(x,L) =\delta(1-x) +\sum_{1}^{\infty} (\frac{\alpha}{2\pi})^n
\frac{1}{n!}(L-1)^n P^{(n)}(x), \\ P^{(n)}(x)=\int_x^1
\frac{dy}{y} P(y)P^{(n-1)}(\frac{x}{y}), n \geq 2.
\end{eqnarray}
In \cite{Fadin} another form of $D(x,L) $ was found
\begin{eqnarray}
D(x,L) =\frac{\beta}{2}[(1-x)^{\frac{\beta}{2}-1}[1+\frac{3}{8}\beta]-\frac{1}{2}(1+x)]+O(\beta^2), \nn \\
\beta=\frac{\alpha}{2\pi}(L-1).
\end{eqnarray}
In Appendix A  we give the expression for the virtual correction to the two quantum
annihilation cross section in the high energy large scattering angle limit.
Its analytic form is given above (see (15)). The result of tabulation of $K_{SV}$ is
given in the table.
The total value of $K_2$ is $K_2=K_{VS}+K^{hard}$.

\section{$3\gamma$ annihilation channel}

For completeness, we also give the cross section of $3\gamma$ annihilation. In \cite{KS95} it was
shown that in the leading logarithmical approximation it has the form (4) with replacement $d\sigma_B$ by $d\sigma^{hard}$
(see(12)). The value  of K-factor ($K=K_3$) is rather a  cumbersome function of  kinematical invariants. Rather realistic estimation
for it can be obtained by using the averaging procedure: replacing the ratio of amplitudes of nonleading contributions to an
amplitude of leading ones and expressing them in terms of cross sections.  The cross section of $3\gamma$ annihilation through the light-by-light
mechanism turns out to be dominant among nonleading terms. This contribution does not contain large logarithms $L$.

This cross section was calculated in \cite{Baier80}
\ba
\sigma^{e^{+}e^{-} \to 3\gamma}_{lbl}(s)=\frac{\alpha^5}{18\pi^2
s}N, \quad  N=200\xi_5 -8\pi^2 \xi_3 +\frac{7}{15} \pi^4 -128 \xi_3
+\frac{41}{3}\pi^2 -124 \approx 15.
\end{eqnarray}

In \cite{Eidelman}, the so called "total" cross section of the process $e^{+}e^{-}
\to 3\gamma$ in the lowest order of PT  was obtained.  The detection energy threshold of final photons
$\frac{2\omega_i}{E}=\nu_i > \eta$. Was implied moreover $\psi_0,
(\cos\psi_0=z\sim 1)$-is the minimal angle between the plane of photons
momentum and beam axis was fixed. It has a form:
\begin{eqnarray}
\sigma_{"tot"}(z)=\frac{2\alpha^3}{s}[\phi_1(z)\ln\eta+\phi_2(z)].
\ea

The Explicit form of functions $\phi_1(z), \phi_2(z)$ is given in Appendix B.

Keeping in mind the  smooth behavior of the non-leading contributions
in the kinematic region of large-angle photon emission, we
estimate  the $K_3$ using the "total cross section" approximation:
\begin{eqnarray}
1+\frac{\alpha}{\pi} K_3 \approx 1+2\sqrt{\frac{\sigma_{lbl}}{\sigma_{"tot"}}},
\quad K_3= \frac{1}{3}\sqrt{N/(\phi_1(z) \ln \eta +
\phi_2(z))}.
\end{eqnarray}
Numerical estimation for $z=0.3$ and $\eta = 0.05$ leads to $1+(\alpha/\pi)K_3\approx 1.01$.

\section*{Acknowledgements}
One of us (EAK) acknowledges the support of  to the Institute of Physics, SAS and INTAS Grant 05-1000008-8328.
The work also was partly supported by the Slovak Grant Agency for
Sciences VEGA, Grant No. 2/7116/27.
We are grateful to Yu. M. Bystritskii for interest in this paper, and G.V. Fedotovich for useful critical remarks.

\section*{Appendix A}
In the case under consideration the result of calculation of the lowest order
correction due to virtual photons  has a form:
\ba
\frac{d\sigma_V}{d\sigma_B}=1+\delta_V,
\ea

\ba
\delta_V=\frac{\alpha}{\pi}[-(L-1)\ln\frac{m^2}{\lambda^2}+\frac{\pi^2}{2}+\frac{1-c^2}{2(1+c^2)}(G(\chi_1,\chi_2)+
G(\chi_2,\chi_1))],\nn
\ea
expression form is rather cumbersome in general case(\cite{Feynman}) $G(\chi_1,\chi_2)$ has a form: In the large
angle kinematic it is
\ba
G(\chi_1,\chi_2)=(1+\frac{s^2}{2\chi_1\chi_2})(\frac{1}{4}L^2-\frac{\pi^2}{12})+[-2L(\frac{\chi_2}{2\chi_1}+\frac{\chi_1}{\chi_2}+1)+ \nn  \\
+\frac{3\chi_2}{2\chi_1}+1]\ln\frac{\chi_1}{m^2}+\frac{1}{4}(L^2-\pi^2)(\frac{7\chi_1}{s}+\frac{3\chi_1^2}{s\chi_2})-L- \nn \\
\frac{3\chi_1}{2\chi_2}+2(\frac{\pi^2}{3}+\frac{1}{2}\ln^2\frac{\chi_1}{m^2})(\frac{\chi_1}{\chi_2}+1+\frac{\chi_2}{2\chi_1}).
\ea
Substituting here $\chi_1=(s/2)(1+c), \chi_2=(s/2)(1-c)$ and taking into account soft photon emission contribution
we arrive after some algebra at expression (\ref{8}).

\section*{Appendix B}
The explicite forms of $\phi_1, \phi_2$ are
\ba
\phi_1(z)= -2\ln(1-z^2)-\frac32
\ln^2(1-z^2)+2\int_0^{z^2}\frac{dx}{x} \ln(1-x), \\ \nonumber
\phi_2(z)=-\frac12\ln^2(1-z^2) +\frac16 \pi^2 \Big(
\frac{2z^2}{1-z^2}-\ln(1-z^2)\Big)+\frac16 \ln^3(1-z^2) \\
\nonumber+ \Big( \frac{2}{1-z^2}+2
\ln\frac{z^2}{1-z^2}\Big)\int_0^{z^2}\frac{dx}{x} \ln(1-x) \\
+\frac32 \int_0^{z^2}\frac{dx}{x} \ln^2(1-x)
-2\int_0^{z^2}\frac{dx}{x} \ln x \ln(1-x).
\ea

\end{document}